\documentclass[iicol,pdflatex,sn-mathphys-num]{sn-jnl}

\usepackage{graphicx}%
\usepackage{multirow}%
\usepackage{amsmath,amssymb,amsfonts}%
\usepackage{amsthm}%
\usepackage{mathrsfs}%
\usepackage[title]{appendix}%
\usepackage{xcolor}%
\usepackage{textcomp}%
\usepackage{manyfoot}%
\usepackage{booktabs}%
\usepackage{algorithm}%
\usepackage{algorithmicx}%
\usepackage{algpseudocode}%
\usepackage{listings}%

\theoremstyle{thmstyleone}%
\theoremstyle{thmstyletwo}%

\theoremstyle{thmstylethree}%

\begin{document}

\title[Article Title]{Disentangling the Effects of Phase-Specific Transport Properties on Elliptic Flow Centrality Dependence in the AMPT Model}

\author*[1]{\fnm{Yao} \sur{Zhang}}\email{yaozhang@jxust.edu.cn}
\author[1]{\fnm{Xipei} \sur{Zhao}}

\affil[1]{\orgdiv{School of Carbon Neutrality}, \orgname{Jiangxi University of Science and Technology}, \city{Nanchang} \postcode{330013}, \country{China}}

\abstract{The centrality dependence of charged hadron elliptic flow $v_{2}$ in Au+Au collisions at $\sqrt{s_{NN}}=200~\rm{GeV}$ is studied using the AMPT model with phase-specific control of partonic and hadronic transport properties. By separately varying the shear viscosity $\eta/s$, phase transition temperature $T_{tr}$, and hadronic scattering cross sections, we isolate their distinct effects on anisotropic flow. Hadronic rescattering is found to contribute increasingly to $v_{2}$ with centrality, with the contribution becoming dominant in peripheral collisions. Scaling up hadronic cross sections by a factor of four substantially amplifies $v_{2}$ in peripheral collisions and improves agreement with STAR data, suggesting that the flattening of $v_{2}$ at high centrality may reflect a larger effective hadronic cross section and highlighting the significant role of the hadronic phase in small systems.}




\maketitle

\section{Introduction}\label{sec1}

Relativistic heavy-ion collisions at the BNL Relativistic Heavy-Ion Collider (RHIC) and the CERN Large Hadron Collider (LHC) provide a unique laboratory for creating and studying the quark-gluon plasma (QGP) which is a deconfined state of strongly interacting matter under extreme temperature and energy density conditions \cite{1,2,3,4,5,6,7}. A primary goal of contemporary high-energy heavy-ion physics is to elucidate the transport properties of the QGP and to pinpoint the critical point of the QCD phase transition. Among the key observables sensitive to the early-stage dynamics and transport coefficients of the produced medium, anisotropic flow, particularly elliptic flow $v_{2}$, which quantifies the azimuthal anisotropy of particle emission, has emerged as a crucial probe. The magnitude and centrality dependence of $v_{2}$ reflect the interplay between the initial geometric anisotropy of the collision overlap region, the subsequent pressure gradients, and the dissipative effects during the system's evolution \cite{8,9,10,11,12}. 

The observation of large elliptic flow at RHIC and LHC, consistent with nearly perfect fluid behavior, has provided strong evidence for the formation of a strongly coupled QGP with a shear viscosity to entropy density ratio $\eta/s$ close to the KSS bound. Specific shear viscosity has been incorporated into viscous relativistic hydrodynamics as an input parameter to extract semi-quantitative constraints from measurements of particle spectra and anisotropic flow \cite{11,12,13,14}. Bayesian parameter estimation methods applied to hybrid hydrodynamics-hadron transport models have further provided precise estimates of the viscosity coefficient and its temperature dependence \cite{15,16,17}. While hydrodynamic models attribute $v_{2}$ trends largely to the specific shear viscosity of the QGP, transport frameworks offer complementary insights by revealing the microscopic mechanisms that drive momentum-space anisotropy \cite{18,19,20}.

The AMPT (A Multi-Phase Transport) model is a self-contained kinetic theory-based description of relativistic nuclear collisions that contains multi-stage structure which provides favorable conditions for disentangling the effects of nuclear matter properties at different evolution stages on both the system evolution and final-state observables \cite{21,22,23}. Previous studies have established that partonic interactions play a dominant role in generating elliptic flow, while hadronic rescattering primarily modifies particle spectra and HBT radii \cite{23}. However, the contribution and magnitude of influence of hadronic rescattering on elliptic flow, as well as on the residual anisotropic flow and momentum redistribution within transport models, remains debated \cite{18,24,25,26}. 

In this work, we present a comprehensive investigation of the centrality dependence of charged hadron elliptic flow in Au+Au collisions at $\sqrt{s_{NN}}=200~\rm{GeV}$ within the AMPT model to separate control mechanisms for the partonic and hadronic transport processes and disentangle the distinct effects of the phase-specific properties on $v_{2}$ and its eccentricity scaling $v_{2}/\varepsilon$. Sec.2 describe the phase-specific transport processes implemented in the AMPT model. Sec.3 presents the results on $v_{2}$ compared to experimental data. Conclusions are summarized in Sec.4.

\section{Phase-Specific Transport Processes in the AMPT Model}\label{sec2}

The AMPT model, as a self-contained hybrid model based on kinetic theory for describing relativistic heavy ion collisions, consists of four components: the fluctuating initial condition, partonic interactions, hadronization, and hadronic interactions, which are realized by the HIJING, ZPC, quark coalescence, and ART models, respectively \cite{18,27,28,29,30}. The evolution stages of each component are relatively independent and the constituent particles evolve discretely, thereby providing favorable conditions for disentangling the effects of nuclear matter properties at different stages on both the system evolution and final-state observables. The cascade method of non-equilibrium dynamics used in the transport processes of partons and hadrons can correlate the scattering cross sections with the phase-specific transport properties.

\subsection{Shear viscous transport dynamics simulation of parton matter}
The partonic transport process in the ZPC parton cascade model with two-body forward-angle scatterings has a natural shear viscosity that depends on the elastic scattering cross section $\sigma_{p}\approx\frac{9\pi\alpha_{s}^{2}}{2\mu^{2}}$ which can be adjusted by changing the strong coupling constant $\alpha_{s}$ and the screening mass $\mu$, and the shear viscosity can be expressed as  \cite{31,32}
\begin{equation}
	\eta=\frac{4T}{5\sigma_{p} g(w)}
\end{equation}
by kinetic theory with the Chapman-Enskog method for anisotropic scattering, where $w=\mu/T$, $g(w)$ is the thermal average of $h(a)=4a(1+a)[(1+2a)\ln(1+\frac{1}{a})-2]$ and can be approximated by
\begin{equation}
	g(w)=h(\frac{w^{2}}{v^{2}}),
\end{equation}
with $v=11.31-4.847\exp(-0.1378w^{0.7338})$. At temperature $T$, the parton matter of massless three-flavor quarks is assumed and the shear viscosity to entropy density ratio is \cite{20,22}
\begin{equation}
	\eta/s\approx\frac{2\pi w^{2}}{45g_{B}\alpha_{s}^{2}g(w)},
\end{equation}
where $g_{B}=52$ when using Boltzmann distribution. The specific shear viscosity $\eta/s$ depends on the ratio of $\mu/T$ for a fixed $\alpha_{s}$ and can be parameterized by adjusting $\mu$ with $T$ in the model.

The space-time distribution of $T$ can be extracted from the discrete partons. For the efficiency of numerical calculations, the volume of the parton space is divided into cells, each with the width of $\Delta x=0.2\rm{fm}$, $\Delta y=0.2\rm{fm}$ and space-time rapidity $\Delta\eta_{\rm{s}}=0.2$, where the temperature at the center of each cell represent the entire cell. The temperature of partonic phase can be estimated with the energy density, $T=(\frac{\pi^{2}\epsilon}{3g_{B}})^{\frac{1}{4}}$. By applying a Gaussian smoothing approximation, the energy of each pointlike parton is convolved with a Gaussian kernel, which allows for obtaining the energy density at evolution time $t$, representable as \cite{33,34}
\begin{equation}
\begin{split}
&\epsilon(x,y,z)=\\
&\sum N_{i}\exp\Biggl[-\frac{(x-x_{i})^{2}+(y-y_{i})^{2}+(z-z_{i})^{2}}{2\sigma_{i}^{2}}\Biggl],
\end{split}
\end{equation}
where $N_{i}=(\frac{1}{2\pi})^{\frac{3}{2}}\frac{1}{\sigma_{i}^{3}}E_{i}$ provides the proper normalisation, $(x_{i},y_{i},z_{i})$ and $E_{i}$ are the position vector and energy of parton $i$, Gaussian width $\sigma_{i}\propto \frac{1}{2E_{i}}$ inversely related to the parton energy is adopted. 

The shear viscous transport dynamics simulation, by utilizing the temperature distribution of parton matter, can achieve the quantitative control of parameterized specific shear viscosity $\eta/s(T)$ and phase transition temperature $T_{tr}$ separately in the model. The scattering cross section $\sigma_{p}$ of partons within a cell is determined by the set $\eta/s$ coefficient and $T$ of the cell at that time, ensuring $\eta/s$ remains consistent with the set value during the evolution of parton system, and is set to zero when the cell temperature is lower than $T_{tr}$, enabling the control of phase transition temperature.

\subsection{Proportional scaling approach of hadronic scattering cross sections}
The hadronic transport process in the ART hadron cascade model includes elastic and inelastic scatterings which exhibit varying hadronic scattering cross sections $\sigma_{H}$ across different hadronic scattering channels. As in partonic process, the hadronic scattering cross sections serve as the most critical parameters in the transport process and are directly related to the transport properties. The occurrence of hadronic scattering collisions is conditioned by their separation distance being smaller than the maximum collision distance, which scales with the square root of $\sigma_{H}$. Hadronic transport model is often employed to provide a microscopic description in the late hadronic phase at low temperature, which overestimates the shear viscosity of hadron matter, especially near the phase transition temperature, the shear viscosity significantly exceeds the value estimated by viscous relativistic hydrodynamic simulations close to the KSS bound in relativistic heavy ion collisions. Considering the inverse relationship between scattering cross section and shear viscosity, and given that many of the hadronic cross sections used in the model have not been thoroughly studied theoretically nor adequately constrained by experimental data, it is reasonable to use a larger $\sigma_{H}$ in transport simulations, which effectively corresponds to a smaller hadronic shear viscosity.

However, unlike $\sigma_{p}$ that can be directly adjusted within the model, the hadronic cascade contains numerous scattering channels with distinct cross sections, making it impossible to globally regulate $\sigma_{H}$ by modifying individual channel cross sections. The proportional scaling approach of hadronic scattering cross sections achieves an equivalent proportional adjustment of all hadronic cross sections by uniformly increasing the maximum collision distance $d_{M}=\sqrt{\sigma_{H}/\pi}$ across all scattering channels. Notably, scaling factors of 2 times in maximum collision distance equate to 4 times amplification in $\sigma_{H}$ \cite{23,26}. Given that $\sigma_{H}$ is not an adjustable free parameter, the scaling of $\sigma_{H}$ in this study is not intended to reflect physical reality but solely serves as a qualitative probe of hadronic rescattering. Future work should incorporate temperature-dependent $\sigma_{H}$ and constrain $\sigma_{H}$ with experimental measurements to better approximate realistic collision dynamics.

The tuning of partonic and hadronic cross sections enables the investigation of phase-specific transport dynamics within the model. To disentangling the effects of phase-specific transport properties, heavy-ion collision events are generated in the AMPT model with different scenarios for the specific shear viscosity, phase transition temperature, and hadronic scattering cross sections for comparison. $\eta/s$ adopts three different parameterizations: constant values of 0.08 and 0.16, and a temperature-dependent $\eta/s(T)$ by a Bayesian estimation with $\eta/s_{\rm{min}}=0.085$, $b=0.83$ and $c=-0.37$, where \cite{15}
\begin{equation}
	\eta/s(T)=\eta/s_{\rm{min}}+b(T-T_{tr})\left(\frac{T}{T_{tr}}\right)^{c}.
\end{equation}
$T_{tr}$ is set to 154, 165 and 175 MeV, and for hadronic cross sections, comparisons are made among the default $\sigma_{H}$, $4\times\sigma_{H}$ and the scenario without hadronic cascade. The parameters of the Lund string fragmentation $a_{L}=0.55$ and $b_{L}=0.2\rm{GeV^{-2}}$, $\alpha_{s}=0.33$ are chosen for Au+Au collisions at $\sqrt{s_{NN}}=200\rm{GeV}$ in the model.

\section{Centrality Dependence of Charged Hadron Elliptic Flow}\label{sec3}

The dependences of various observables on the size or centrality of the collision system are useful for exhibiting the onset or transition of certain phenomena, such as momentum anisotropy from initial collision geometry or final-state interactions. The system size dependence of anisotropic flow can provide key information on the origin of collectivity and the transport properties of the system evolution. For small colliding systems in particular, transport models based on non-equilibrium dynamics may differ markedly from hydrodynamic descriptions.

\begin{figure}[h]
	\centering
	\includegraphics[width=0.5\textwidth]{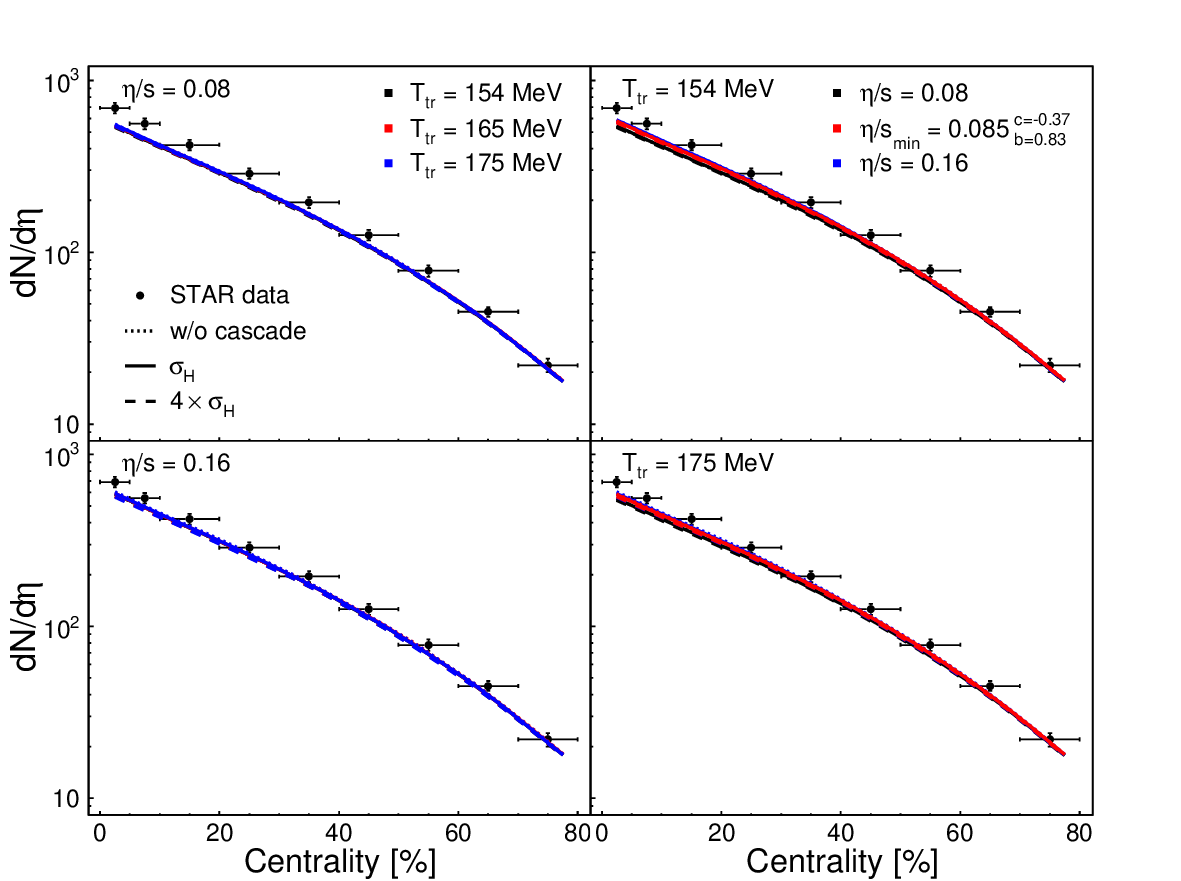}
	\caption{Centrality dependence of charged hadron multiplicity for Au+Au collisions at $\sqrt{s_{NN}}=200~\rm{GeV}$ with various $\eta/s$, $T_{tr}$ and $\sigma_{H}$ compared to experimental data from the STAR collaboration \cite{22}.}\label{fig1}
\end{figure}

Figure 1 shows the mid-pseudorapidity multiplicity of charged hadrons versus centrality for Au+Au collisions at $\sqrt{s_{NN}}=200~\rm{GeV}$ with various $\eta/s$, $T_{tr}$ and $\sigma_{H}$ compared to STAR data. It is seen that only $\eta/s$ has a weak increasing effect on $dN/d\eta$, while $T_{tr}$ and $\sigma_{H}$ have negligible influence on the results. The slight increase in mid-pseudorapidity hadron yields due to shear viscosity suppressing longitudinal expansion cannot fill the gap between the model and experiment, especially in central collisions where the results are significantly lower than the experimental data. It has recently been found that using the local nuclear scaling approach to treat the Lund parameter $b_{L}$ and the minijet cutoff $p_{0}$ in the HIJING initial condition as local variables that depend on the local nuclear thickness functions from the two colliding nuclei in each event can improve the system size and centrality dependences of the model, yielding mid-pseudorapidity multiplicity consistent with experimental results \cite{19,36}.

Anisotropic flow, which can be characterized by flow harmonics of the particle azimuthal angle distribution, originates from pressure gradients induced by the initial geometry of the collision and is sensitive to the transport properties and processes of the system. The second harmonic coefficient, elliptic flow $v_{2}$, represents a key measurable quantity and can be calculated by the two-particle cumulant method as
\begin{equation}
	v_{2}\{2\}=\sqrt{\langle\mathrm{cos}2(\phi_{1}-\phi_{2})\rangle},
\end{equation}
with $\langle...\rangle$ denotes an average over all particle pairs in each event and then over all events \cite{37}.

Figure 2 and figure 3 show the centrality dependence of charged hadron elliptic flow $v_{2}$ with various $\eta/s$, $T_{tr}$ and $\sigma_{H}$ compared to STAR data. While it is unsurprising that $\eta/s$, $T_{tr}$ and $\sigma_{H}$ all have significant effects on elliptic flow, the differential behaviors of phase-specific transport processes in distinct centrality regions reveals intriguing results. Comparing the curves without hadronic cascade, it is evident that both enhanced shear viscosity and increased phase transition temperature lead to notable suppression of elliptic flow, which remains visible after hadronic rescattering. Increasing $\eta/s$ impedes the conversion from spatial anisotropy to momentum anisotropy and reduces the buildup of collective flow, while raising $T_{tr}$ shortens the partonic phase evolution, thereby reducing the conversion time and consequently weakening the flow values. As the collision becomes more peripheral, the difference between the final-state elliptic flow obtained with hadronic rescattering and that without hadronic cascade gradually enlarges from a small deviation in central collisions to a substantial one, with the discrepancy becoming significant in peripheral collisions, indicating that the contribution of hadronic rescattering to $v_{2}$ grows with increasing centrality. 

\begin{figure}[h]
	\centering
	\includegraphics[width=0.5\textwidth]{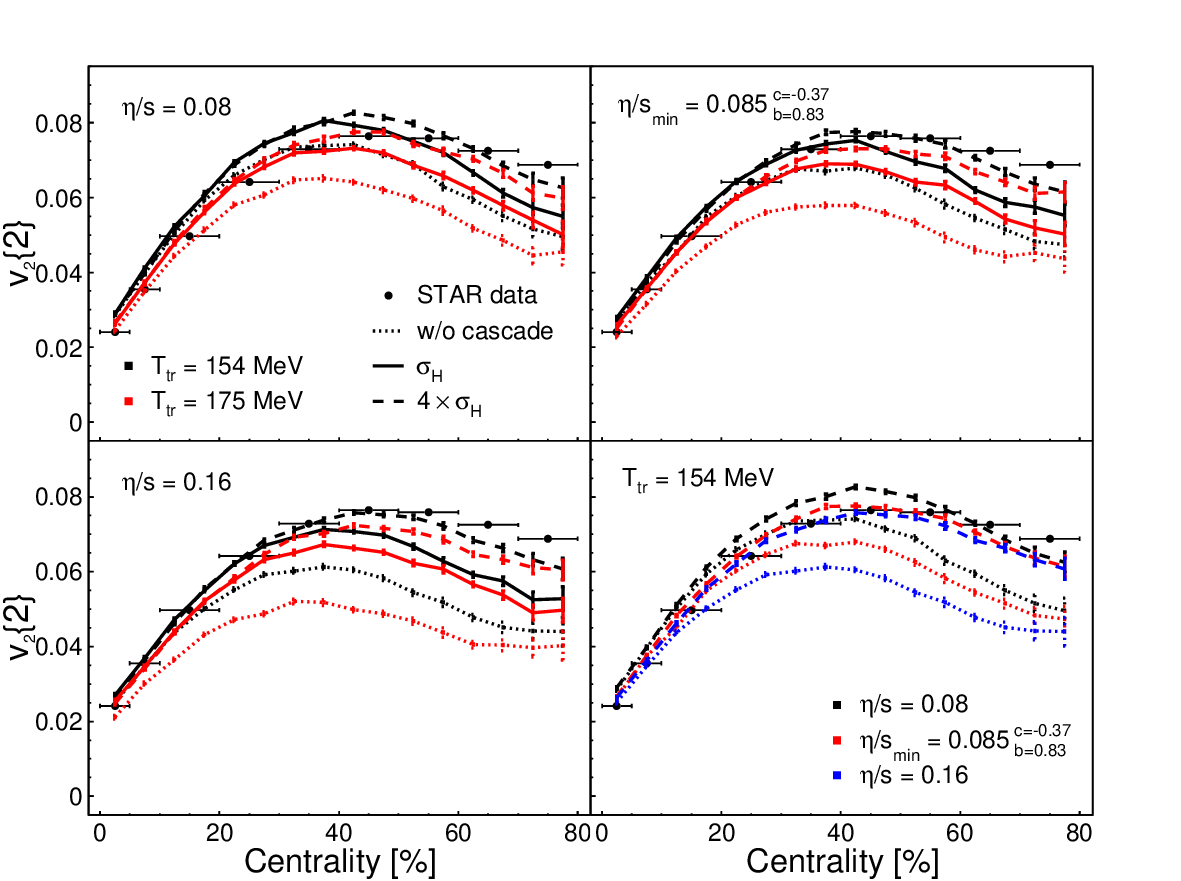}
	\caption{Centrality dependence of charged hadron elliptic flow for Au+Au collisions at $\sqrt{s_{NN}}=200~\rm{GeV}$ with various $\eta/s$ and $\sigma_{H}$ compared to experimental data from the STAR collaboration \cite{38}. Errors are statistical only.}\label{fig2}
\end{figure}

\begin{figure}[h]
	\centering
	\includegraphics[width=0.5\textwidth]{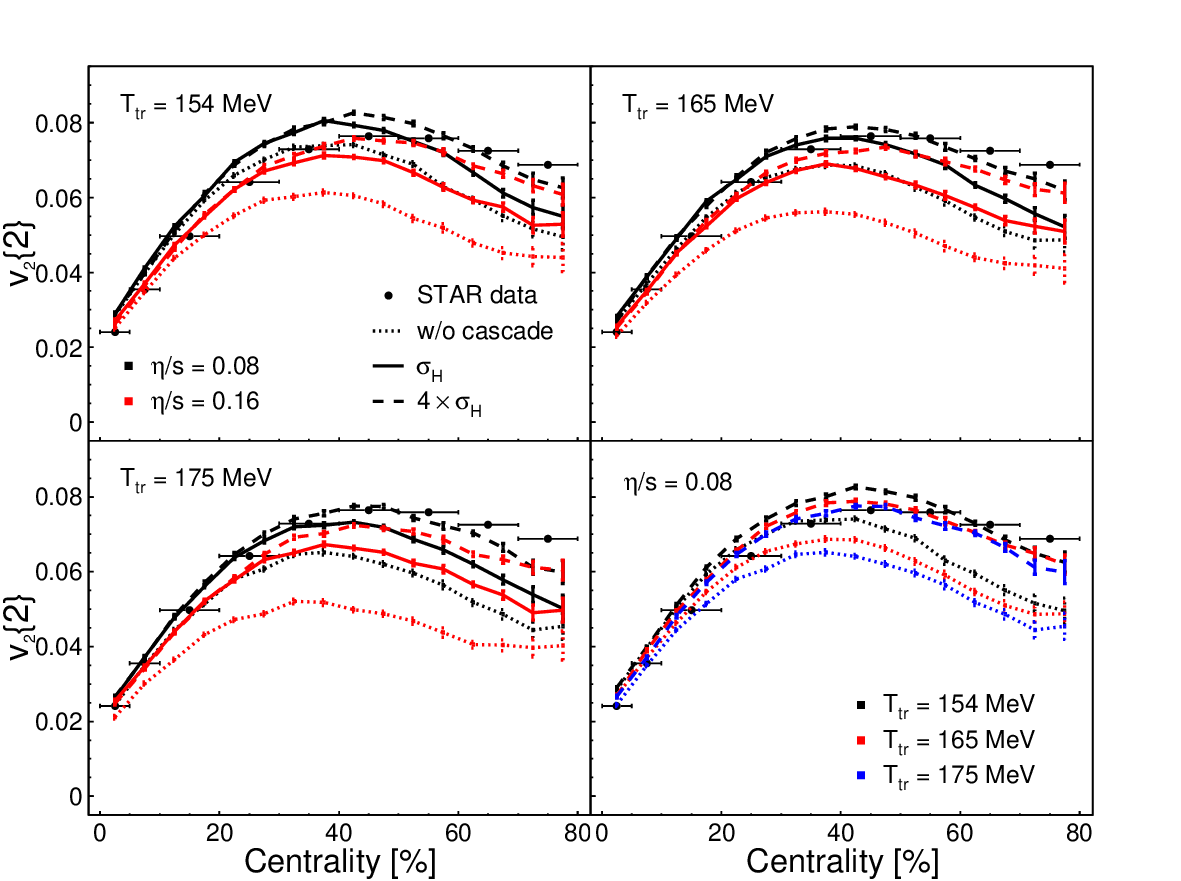}
	\caption{Same as Fig.2 but with various $T_{tr}$ and $\sigma_{H}$.}\label{fig3}
\end{figure}

A comparison of the curves for different $\eta/s$ and $T_{tr}$ shows that the contribution of hadronic rescattering to $v_{2}$ is further enhanced with increasing $\eta/s$ or $T_{tr}$. The centrality dependence of hadronic rescattering contribution on $v_{2}$, linked to $\eta/s$ and $T_{tr}$, arises from the anisotropic evolution of the system. Elevated collision centrality enhances the initial spatial anisotropy and shortens the partonic phase evolution, resulting in amplified residual spatial anisotropy within the hadronic system that is accessible for conversion via hadronic rescattering. Meanwhile, increasing $\eta/s$ or $T_{tr}$ both impede the partonic phase anisotropy conversion from coordinate to momentum space and leave larger residual coordinate-space anisotropy in the hadronic phase after hadronization. This enhanced residual anisotropy enables hadronic rescattering to generate a larger momentum anisotropy. Moreover, when the hadronic scattering cross section is increased to four times the default value, the larger cross section $4\times\sigma_{H}$ is found to once again significantly amplify $v_{2}$ in most peripheral collisions, leading to better agreement with experiments. This suggests that the flattening of the centrality dependence of elliptic flow at high centrality may correspond to a larger hadronic scattering cross section and the contribution of hadronic rescattering to elliptic flow is surprisingly important in small colliding systems. In addition to the constant specific shear viscosity, the elliptic flow results using the temperature-dependent $\eta/s(T)$ by a Bayesian estimation are also compared. The value of $\eta/s(T)$ decreases with falling temperature and tends toward 0.085. As seen in the figure, $v_{2}$ with $\eta/s(T)$ in central collision is closer to the result from the constant $\eta/s=0.16$. With increasing centrality, the system evolution time shortens, and $v_{2}$ with $\eta/s(T)$ gradually become closer to the result from the constant $\eta/s=0.08$. Compared to the constant $\eta/s$, the temperature-dependent specific shear viscosity yields a more diverse elliptic flow as a function of centrality, which allows for improved agreement with the experimental data.

\begin{figure}[h]
	\centering
	\includegraphics[width=0.5\textwidth]{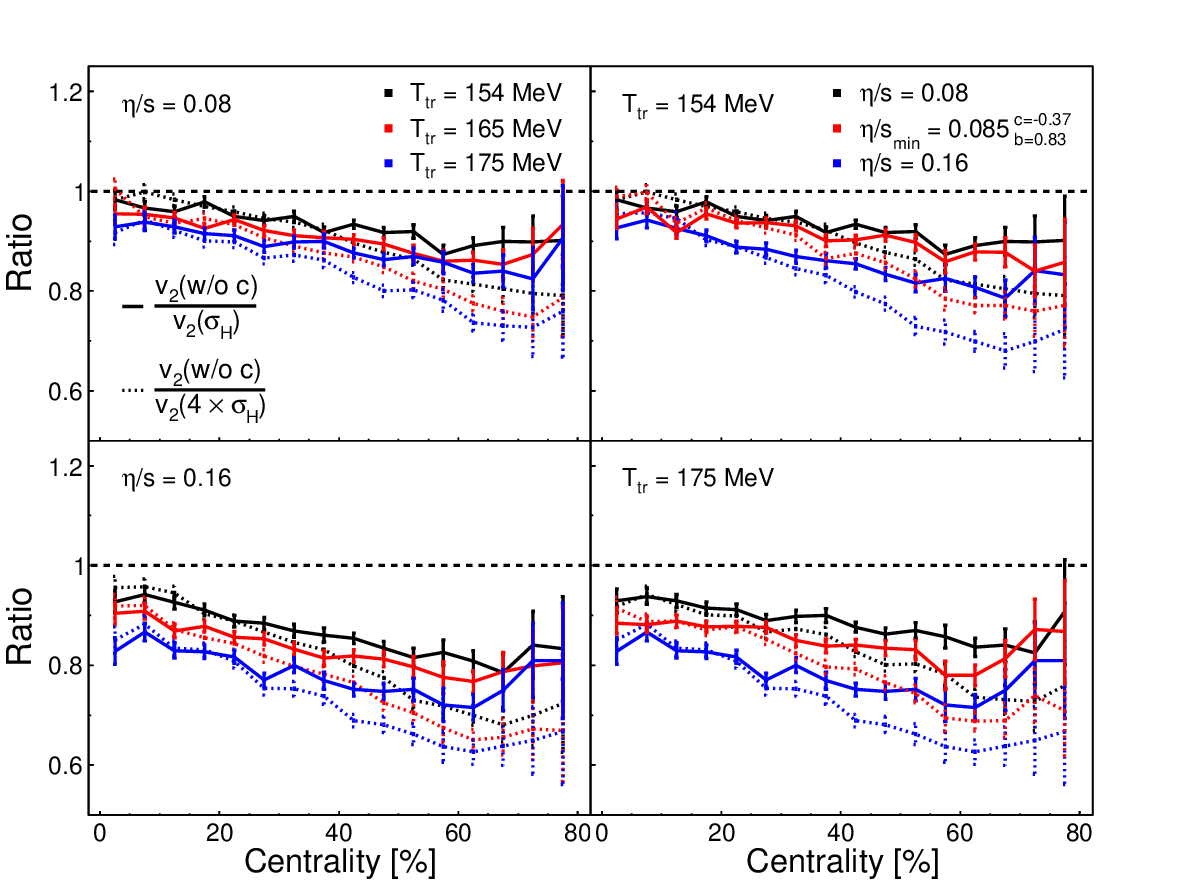}
	\caption{Ratio of $v_{2}$ without hadronic cascade to that with  $\sigma_{H}$ and with $4\times\sigma_{H}$, respectively, as a function of centrality.}\label{fig4}
\end{figure}

To illustrate in more detail the centrality dependence of the fractional contribution of hadronic rescattering to elliptic flow, Figure 4 shows the ratios of $v_{2}$ without hadronic cascade to that with  $\sigma_{H}$ and with $4\times\sigma_{H}$ cross sections, respectively. A smaller ratio indicates a larger fractional contribution from hadronic rescattering. Consistent with the analysis of the centrality dependence of $v_{2}$, the ratio curves shown in the figure decline as centrality increases with a roughly stable slope, indicating that the fractional contribution of hadronic rescattering to $v_{2}$ increases with centrality. It is noteworthy that the ratio curve with $4\times\sigma_{H}$ exhibits a significantly faster decline than that with $\sigma_{H}$. This decline is substantially larger than that driven by the $v_{2}$ suppression of partonic phase from increased $\eta/s$. The phase transition temperature, however, mainly alters the upper limit of the ratio with only a minor influence on the slope of the curve. Raising $T_{tr}$ tends to shift the curve downward. 

Analysis of the elliptic flow results reveals that $\eta/s$, $T_{tr}$ and $\sigma_{H}$ exhibit distinct and separable effects on the centrality dependence of anisotropic flow. Anisotropic flow originates from the conversion of initial spatial anisotropy into momentum anisotropy through anisotropic pressure gradients caused by the very strong interactions among the initially produced partons. Its magnitude is strongly correlated with the initial anisotropic shape of the collision and should be proportional to the initial eccentricity $\varepsilon$ in the transverse plane. Consequently, it is also necessary to compare the effects of phase-specific transport properties on the eccentricity-scaled elliptic flow $v_{2}/\varepsilon$.

The spatial asymmetry in the transverse plane is commonly given by the standard eccentricity $\varepsilon_{std}=\frac{\langle{y^{2}-x^{2}}\rangle}{\langle{y^{2}+x^{2}}\rangle}$, with $x$ and $y$ are the position of particles. Because of initial fluctuations, the appropriate eccentricity $\langle\varepsilon_{part}\rangle$ and $\varepsilon_{part}\{2\}$ can be expressed as \cite{39}
\begin{equation}
	\varepsilon_{part}=\frac{\sqrt{(\sigma_{y}^{2}-\sigma_{x}^{2})^{2}+4\sigma_{xy}^{2}}}{\sigma_{y}^{2}+\sigma_{x}^{2}}
\end{equation}
and
\begin{equation}
	\varepsilon_{part}\{2\}=\sqrt{\langle\varepsilon_{part}^{2}\rangle},
\end{equation}
where $\sigma_{x}^{2}=\{x^{2}\}-\{x\}^{2}$, $\sigma_{y}^{2}=\{y^{2}\}-\{y\}^{2}$ and $\sigma_{xy}=\{xy\}-\{x\}\{y\}$ are the variances of particle distribution, while $\langle...\rangle$ and $\{...\}$ represent the averaging over all events and over all particles in a single event, respectively. The initail ellipse area $S$ in the transverse plane can be given by
\begin{equation}
	S=\pi\sqrt{\sigma_{x}^{2}\sigma_{y}^{2}-\sigma_{xy}^{2}}.
\end{equation}

The eccentricity $\varepsilon_{part}$ is usually computed for each event from the transverse position distribution of participant nucleons obtained from the Monte Carlo Glauber model. The fluctuating initial condition generated by the HIJING model includes not only the participant nucleon distribution based on the Monte Carlo Glauber framework but also the primordial parton distribution produced by the fragmentation of excited strings in the Lund string model, which allows the model to compute $\varepsilon_{part}$ for both the overlap region of the colliding nuclei and the transverse plane of the initial partons. Figure 5 shows the initial eccentricities $\varepsilon_{std}$, $\langle\varepsilon_{part}\rangle$, $\varepsilon_{part}\{2\}$ and transverse area $S$ as a function of centrality, calculated respectively from the participant nucleons in the MC-Glauber model or from the initial partons at midrapidity. The trends of eccentricities and transverse area with centrality for the initial partonic system are generally in agreement with the results from the nuclear overlap region, but the larger eccentricities, smaller transverse area, and distinct slopes in their centrality dependencies indicate that the shape of the initial partonic system differs from that of the nuclear overlap region.

\begin{figure}[h]
	\centering
	\includegraphics[width=0.35\textwidth]{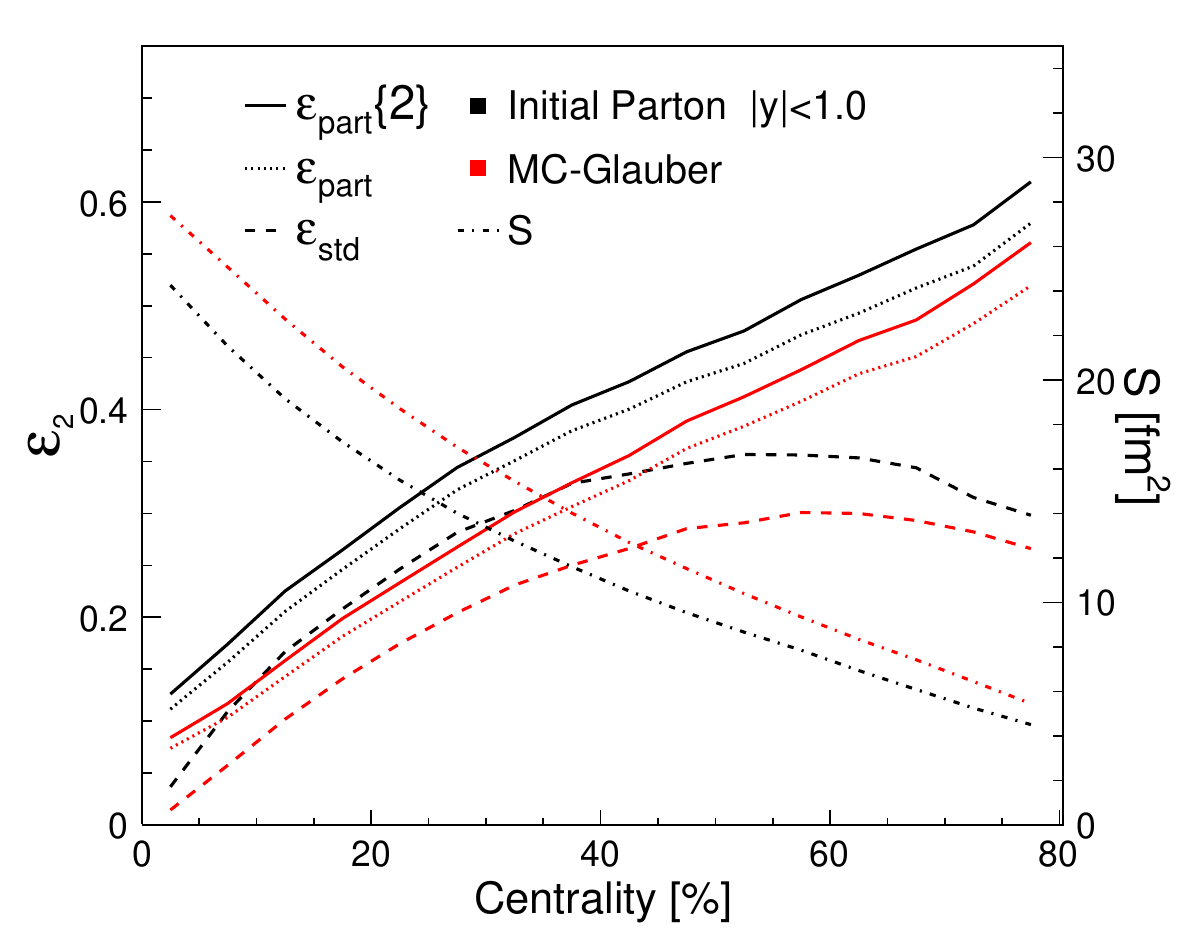}
	\caption{Centrality dependence of the initial eccentricities $\varepsilon_{std}$, $\langle\varepsilon_{part}\rangle$, $\varepsilon_{part}\{2\}$ and transverse area $S$ calculated from the MC-Glauber or the initial distribution of partons, respectively.}\label{fig5}
\end{figure}

\begin{figure}[h]
	\centering
	\includegraphics[width=0.5\textwidth]{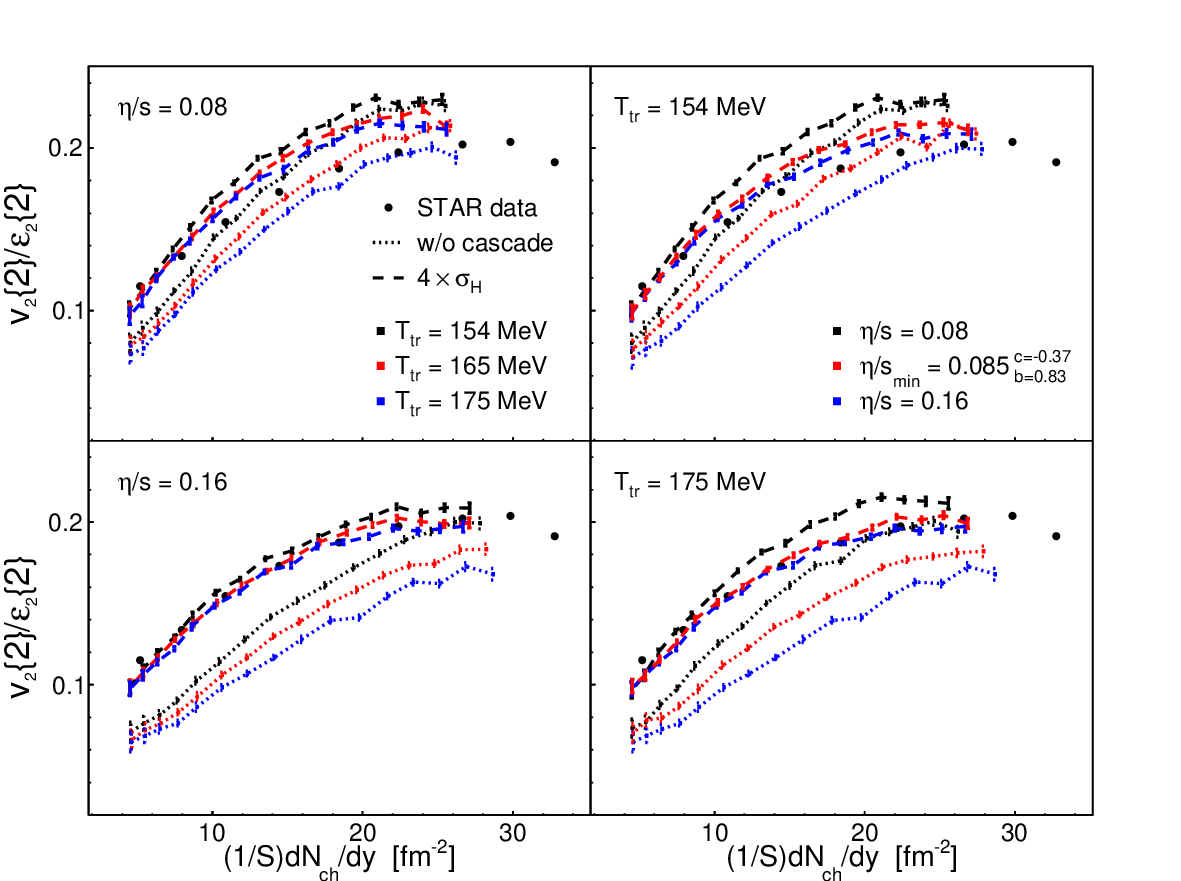}
	\caption{Eccentricity-scaled elliptic flow $v_{2}/\varepsilon$ as a function of multiplicity density $\frac{1}{S}\frac{dN_{ch}}{dy}$.}\label{fig6}
\end{figure}

The eccentricity-scaled elliptic flow $v_{2}/\varepsilon$ is expected to scale with the transverse charged particle area density at mid-rapidity according to $\frac{v_{2}}{\varepsilon}\propto\frac{1}{S}\frac{dN_{ch}}{dy}$. The two-particle cumulant $v_{2}\{2\}$, which naturally includes the contribution of each event fluctuations, should be scaled with the second cumulant $\varepsilon_{part}\{2\}$. Since the fluctuating initial condition, as a component of the model, can provide the initial parton distribution for each event, the calculations of eccentricity and transverse area are replaced from the usual participant nucleons to the initial partons of the collision system. In figure 6, $v_{2}/\varepsilon$ shows a smooth rising trend with increasing transverse density $\frac{1}{S}\frac{dN_{ch}}{dy}$, approaching saturation. The hadronic rescattering with the $4\times\sigma_{H}$ cross section markedly increases $v_{2}/\varepsilon$ in $\frac{1}{S}\frac{dN_{ch}}{dy}\leq20\rm{fm}^{-2}$, flattening the curve with a decreased slope which becomes even more pronounced when $\eta/s$ is enhanced and the resulting curve agrees well with the experimental data. However, it should be emphasized that the curve shifts significantly leftward in the high $\frac{1}{S}\frac{dN_{ch}}{dy}$ corresponding to central collisions, deviating from the experimental data, because the AMPT model underestimates the charged hadron multiplicity in Fig.1 and can be resolved in future work by using the local nuclear scaling of the initial condition parameters. The enhancement of both the final-state elliptic flow and the eccentricity-scaled $v_{2}$ by increasing $\sigma_{H}$ grows with increasing centrality. Interestingly, the centrality region where the suppression of these two observables by enhanced $\eta/s$ or increased $T_{tr}$ is most pronounced shifts from the mid-centrality region for the centrality dependence of $v_{2}$ to the low-centrality region after eccentricity scaling. This not only amplifies the otherwise weak effects of $\eta/s$ and $T_{tr}$ in central collisions, but also provides a possible means to disentangle the effects of the transport properties of the partonic and hadronic phases. 

The effects of $\eta/s$ and $T_{tr}$ on $v_{2}/\varepsilon$ at low $\frac{1}{S}\frac{dN_{ch}}{dy}$ are weak, and are almost negligible especially in the results with hadronic cascade. The suppression of $v_{2}/\varepsilon$ by increased $\eta/s$ and $T_{tr}$ grows with increasing transverse density, and is especially significant for the results without hadronic cascade. The influences of $\eta/s$ and $T_{tr}$ (which determine partonic transport) on anisotropic flow were found to be somewhat similar and hard to tell apart in the earlier findings. However, by comparing the curves with and without hadronic cascade, a separation between their effects becomes visible. Overall, $\sigma_{H}$ and $\eta/s$ control $v_{2}/\varepsilon$ at low and high transverse density: a larger $\sigma_{H}$ gives a larger $v_{2}/\varepsilon$ at low transverse density with a flatter slope, while a smaller $\eta/s$ gives a larger $v_{2}/\varepsilon$ at high transverse density with a steeper slope. As a result, the slope of the curve with hadronic scattering is mainly governed by $\sigma_{H}$ and $\eta/s$. It is important to stress that $T_{tr}$ governs the slope correlation between the curves with and without hadronic scattering: the higher $T_{tr}$, the slopes of the two curves become more similar and their trends align more closely. Thus, the effects of $\sigma_{H}$, $\eta/s$ and $T_{tr}$ of different phases on $v_{2}$ and $v_{2}/\varepsilon$ are effectively separated.

\section{Summary}\label{sec4}

We have systematically studied the phase-specific transport properties in the AMPT model by separately tuning the specific shear viscosity $\eta/s$, the phase transition temperature $T_{tr}$, and the hadronic scattering cross sections $\sigma_{H}$ via a proportional scaling approach. The centrality dependence of charged hadron elliptic flow $v_{2}$ in Au+Au collisions at $\sqrt{s_{NN}}=200~\rm{GeV}$ is analyzed and compared to STAR data. Our results demonstrate that $\eta/s$, $T_{tr}$, and $\sigma_{H}$ each exert distinct and separable influences on $v_{2}$. Increasing either $\eta/s$ or $T_{tr}$ suppresses the partonic conversion of spatial anisotropy into momentum anisotropy, thereby reducing the final $v_{2}$. In contrast, enhancing $\sigma_{H}$ by a factor of four substantially amplifies $v_{2}$ in peripheral collisions, bringing the model into better agreement with experimental data. Notably, the fractional contribution of hadronic rescattering to $v_{2}$ grows steadily with centrality, and this growth is significantly accelerated by larger hadronic cross sections, while $T_{tr}$ mainly affects the overall magnitude rather than the slope of this centrality dependence. After scaling $v_{2}\{2\}$ by the initial eccentricity $\varepsilon_{part}\{2\}$ obtained from the fluctuating initial parton distribution, the phase-specific effects remain clearly distinguishable. The study highlights that hadronic rescattering plays a surprisingly important role in small collision systems, and that a realistic description of the late-stage hadronic transport, especially with enhanced cross sections, is essential for reproducing the flattening of the centrality dependence of elliptic flow. 

These findings underscore the necessity of constraining phase-dependent transport coefficients with experimental observables to achieve a more complete understanding of the evolution dynamics in relativistic heavy-ion collisions. This indicates that the determination of transport properties using elliptic flow $v_{2}$ and its eccentricity scaling $v_{2}/\varepsilon$ should include contributions from different evolution stages. Moreover, a global Bayesian analysis that includes the eccentricity-scaled elliptic flow is also promising and deserves future exploration. 

\backmatter

\bmhead{Acknowledgements}

This work was supported by the Jiangxi Provincial Key Laboratory of Particle Technology (No. 20242BCC32119) and the Supercomputing Platform of Jiangxi University of Science and Technology.

\end{document}